# AN ADAPTIVE DIFFSERV APPROACH TO SUPPORT QOS IN NETWORK MOBILITY NEMO ENVIRONMENT

Loay F. Hussein[1,2], Aisha-Hassan AbdallaHashim[2], Mohamed Hadi Habaebi[2], and Wan Haslina Hassan[3]

[1] Department of Computer and Information Sciences, Jouf University, Saudi Arabia
[2] Department of Electrical and Computer Engineering, Kulliyyah of Engineering, International Islamic University Malaysia, Kuala Lumpur, Malaysia
[3] Department of Electronic Systems Engineering, Universiti Teknologi Malaysia, Kuala Lumpur, Malaysia

*ABSTRACT*

*Network Mobility Basic Support (NEMO BS) protocol (RFC 3963) is an extension of Mobile IPv6. The NEMO BS embraced by IETF working group to permit any node in the portable network to be accessible to the Internet despite the fact the network itself is roaming. This protocol likewise Mobile IPv6 doesn't deliver any kind of Quality of Service (QoS) guarantees to its clients. It can barely offer the same level of services (i.e. Best-Effort) to all the users without obligation to the application's needs. This propositions a challenge to real-time applications that demand a precise level of QoS pledge. The Differentiated Services has recently come to be the most widely used QoS support technology in IP networks due to its relative simplicity and scalability benefits. This paper proposes a new scheme to provide QoS to mobile network nodes within NEMO context. The proposed scheme intends to reduce handover latency for the users of MNN as well as alleviates packet losses. The feasibility of the proposed enhancement is assessed by measuring its performance against the native NEMO BS standard protocol using NS-2 simulator. The obtained results in the simulation study have demonstrated that the proposed scheme outperforms the standard NEMO BS protocol.*

## 1. INTRODUCTION

At the beginning of the twenty-first century, the Internet was fated to be worldwide communication infrastructure. Indeed, even though the Internet presently runs quicker and is expanding in scale, its fundamental design remains unaltered from its initial existence. It basically still operates as a datagram network, where each packet is delivered individually through the network. Delivery time of packets is not ensured, besides packets possibly will be dropped due to overcrowding networks. This impulsiveness does not coordinate well with real-times applications. With the appearance of new applications such as, video-on-demand, voice-over-IP, video conferencing and Internet telephony, the adaptation to serve applications with acceptable Quality of Service (QoS) is required because they cannot withstand transmission data loss or delay jitter. Apparently, E-commerce is revolutionizing the way we do business. QoS is a general term, which means many things to many people. Hardware and software vendors, consumers, researchers, telecommunications operators and so on, seem to have their own definitions of QoS but there is no common definition. In common, the capability to supply resource guarantee and service distinction in any network is frequently alluded to as QoS.

The Internet Engineering Task Force (IETF) has industrialized new standards to grant service distinction and resource guarantee in the Internet, within the sphere term quality of service. The finest recognized QoS methods are Integrated Services (IntServ), Differentiated Service (DiffServ) and Multiprotocol Label Switching (MPLS).

The Integrated service framework changes the elementary IP model in order to accommodate real-time applications and best-effort streams as well. In other words, it was the first attempt to enhance the Internet with QoS capabilities. IntServ is achieved by performing admission control and installing a per-flow state along the path. So, to receive resource assurance from this kind of network, applications must set up the resource reservation along the path before they can start to transmit traffic onto the network. Basically, a resource reservation involves a couple of steps. At the beginning, the application obligates to describe its traffic source and the recourse prerequisites. Afterward, the network utilizes a routing protocol to discover a path based on the demanded assets. Resource reservation protocol known as (RSVP) is used to establish the reservation condition down that path. Only if sufficient assets are available, the admission control welcomes new reservation at any hop. As soon as the reservation is built up, the application can exclusively begin to send traffic across the path. Resource arrangement has to be constrained by packet sorting and scheduling mechanisms in every router in IntServ domain. The IntServ architecture has proposed two classes of service, guaranteed service (RFC 2212) and controlled-load service (RFC 2211). Both of them are based on quantitative service requirements and they require signalling and admission control in each network node[1].

Differentiated Services (DiffServ) architecture is designed to scale to large networks and large customer population. It applies amalgamation of edge patrolling, provisioning and traffic prioritization to accomplish service distinction. The main idea of DiffServ is to push the complexity at the edge of the network, whereas the core of the network is maintained as simple as possible. The Traffic of users is split into a tiny amount of transmission classes. The quantity of traffic that consumers can inject into the network for each forwarding class is restricted at the edge of the network. By altering the complete quantity of traffic permitted in the network, service suppliers can adjust the level of resource supply and therefore regulate the degree of resource assurance for the customer. DiffServ architecture has proposed two classes of services Assured Forwarding (RFC 2597) and Expedited Forwarding (RFC2598) [2].

Finally, Multi-Protocol Label Switching is an advanced forwarding scheme that extends routing with respect to packet forwarding and path controlling. The method used by MPLS is called label switching. The packet header is encoded with a brief fixed-length label and manipulated for packet forwarding. Use the incoming tick in the packet header to discover the next hop and the matching outgoing label when a label switch router (LSR) gets a labelled packet. It also offers per-flow guarantees but encounters some complex issues as MPLS domain routers need to operate distinct routing algorithms to discover the best QoS routes [3].

DiffServ is one of the most outstanding approaches we have listed in this section owing to its simplicity and scalability benefits. Nevertheless, this approach was initially designed without mobility in-mind. Therefore, it has still not been entirely improved to mobile networks, especially in a topology that travels as one piece and tie to random spots in the routing substructure. It seems necessary to integrate QoS with mobility assistance to meet users ' desires.

Due to its significance in military and vehicular applications, NEMO has been famous topics in recent years. Conceptually, the NEMO basic support protocol is simple. A special node Mobile Router (MR) acts both as Mobile IP Mobile Node (MN) as well as a router for the network it serves. Mobile networks encompass at least one mobile router serving them. A Mobile Router at its attachment point (i.e. in the visited network) does not allocate the Mobile Network paths to the infrastructure. Alternatively, it holds a bi-directional tunnel to a home agent (HA) advertising a mobile network accumulation to the establishment.

The rest of the paper is structured as follows. Section 2 will deliberate a background of mobility in general and network mobility basic support protocol in particular. Next, section 3 will explore the related works that gather the integration of QoS within NEMO area. After that, section 4 is devoted to present the proposed scheme and its operations. The performance evaluation and results of the simulation are considered in Section 5. Finally, the conclusion is drawn in section 6.

## 2. MOBILITY BACKGROUND

This section will start by providinga general idea of mobility term and then examining Network Mobility Basic Support Protocol in detail. There are different types of mobility: Host and Network mobility. Host mobility is known also as terminal mobility. It can be referred to as an end host that adjusts its point of attachment to the networks whereas uninterrupted communication between the host and its corresponding node remains. To sustain the integrity of the communication, precise location and routing processes are needed. Mobile IPv6 (RFC 3775, June 2004) [4], Fast Mobile IPv6 (RFC 5268, June 2008) [5], Hierarchical Mobile IPv6 (RFC 5380, October 2008) [6] and Fast Hierarchical Mobile IPv6 [7] are examples of host mobility protocols. On the other hand, Network mobility refers to an entire mobile IP subnet changing dynamically its point of attachment to the Internet backbone;besides its topology accessibility deprived of disturbing the delivery of IP packets to/from the mobile network. A mobile network is usually provided with a mobile router and arrange of mobile nodes in a straightforward network mobility situation. In a complicated mobility situation, mobile nodes or other mobile networks can visit a mobile network themselves. Instances of mobile network protocols are NEMO BS, Nestled NEMO and Multi-homed network as revealed in Figure 1.

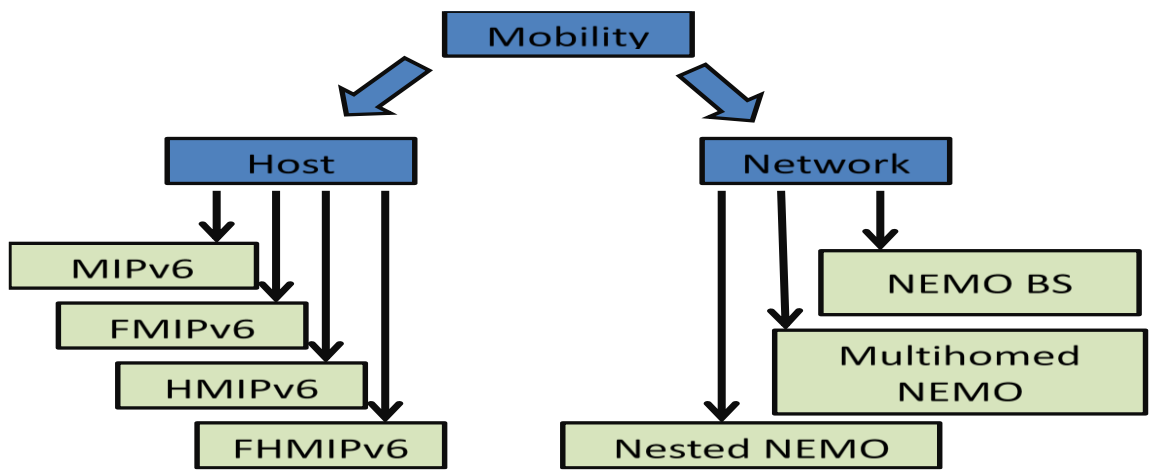

Figure1. Mobility classification

A mobile network (also known as NEMO Basic Support protocol) is described as a network that its Internet point of attachment fluctuates momentarily. In the NEMO architecture, Mobile Router (MR) gains the control of mobility functionality from all the nodes (i.e. MNN), to allow the transparency of mobile platform Internet access. Thus, the nodes inside the mobile network are completely unaware of their network's mobility. Figure 2 shows the reference model of mobility network architecture. Mobile router (MR) functions as a gateway for mobile network nodes. Each node is called Mobile Network Node (MNN).

There are different types of MNNs:

- **Local Fixed Node (LFN):** this stationary node does not roam from a mobile network point of view. Though, it is supported by the MR to achieve connectivity. Hence, its IPv6 address is drawn from an attached NEMO MNP.

- **Local Mobile Node (LMN):** this node often resides in the mobile network. It can also move to other networks. The home network of LMN is situated on the mobile network. In addition, an MNP extracts its home address (HoA).

- **Visiting Mobile Node (VMN):** This node is temporarily connected to the mobile home network from another mobile network. It actually doesn't belong to the mobile Home network and its CoA is taken from an MNP.

The mobile network might include one or more mobile routers that connect it to the global Internet. Each MR must have a Home Agent (HA). The HA is informed of the position of the MR and deflects packets that were addressed to MNNs by the Correspondent Node (CN) and vice-versa. This is the basic requirement for supporting network mobility. A bidirectional tunnelling between MR and HA helps in preserving the continuity of the session while the MR moves. A Mobile Network can also comprise nested subnets. In particular, this is happening when the Mobile Router is attached to another Mobile Network. Namely, Nested NEMO occurs when the VMN is MR and it has its own MNNs. The MR connected straightforwardly to the wired network via an access router (AR) is named the root-MR.

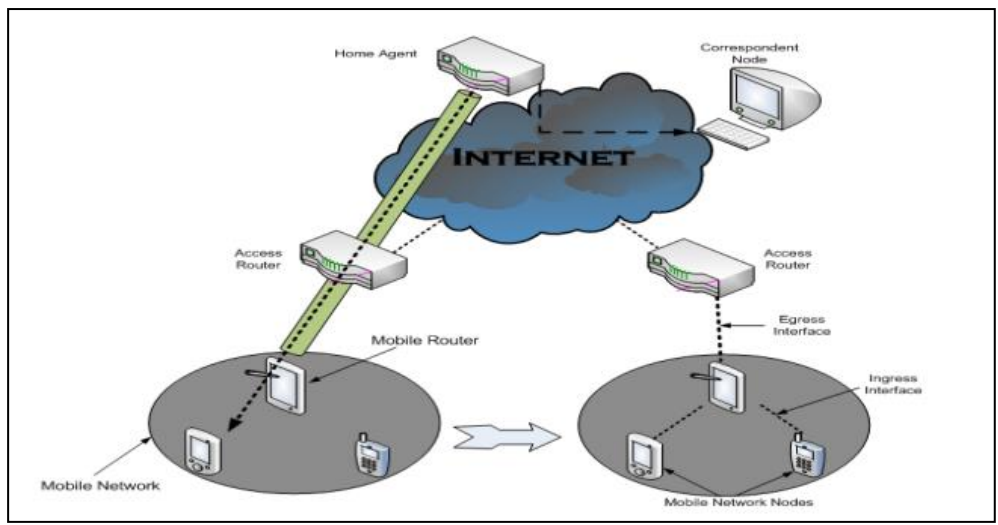

Figure 2. NEMO reference model

The process operation of NEMO BS protocol is similar to Mobile IPv6 for a single host. However, NEMO slightly extends additional mechanisms to support the movement of complete networks.

The Mobile Router (MR) has a distinctive Home Address known as the Mobile Network Prefix (MNP) by which it can be reached when recorded with its Home Agent. The MNP is meant to all IPv6 subnet prefix(es) of the mobile network. MR can possess numerous home addresses if the home link contains various prefixes. So, it can simply advertise one or more of its delegated IPv6 prefixes in the Mobile Network attached to it. All MNN should have an address enclosing this prefix. The encapsulation assists to preserve mobility transparency. Namely, both of the MNN and the correspondent node are totally unaware of the mobility within NEMO while still maintaining the established Internet connections of the MNN. The mobile router receives the encapsulated IP datagram and then encapsulates the tunnel by removing the outer IPv6 header. Afterward, it delivers the destined original datagram to MNN of the NEMO subnet. In the opposite direction, the process is analogous. The mobile router encapsulates the IP datagrams that have been sent by MNN toward its Home Agent, which then decapsulates and forwards the original datagram toward the CN. Even though NEMO method ensures session continuity for every node in the mobile network, it introduces a serious amount of header overhead and builds tunnels on suboptimal routes. This redundant route wastes network resources and increases communication delay[8].

## 3. RELATED WORK

NEMO Basic Support has been approved and published by Internet Engineering Task Force (IETF) in couples of respective Request for Comments (RFC) such as RFC 3963, RFC 4885, RFC 4886, RFC 4887, RFC 4888, RFC 4889, RFC 5177, RFC 4980, RFC 5488, and RFC 6089. Several novel real-life demonstrations [9], [10], [11], [12], [13] and test beds [14], [15], [16], [17],[18] have been working on verifying its feasibility and usability on the Internet. To the best of our knowledge, quite a numbers of research efforts have implemented in the literature to enhance the performance of NEMO BSP by overcoming some issues such as protocol header overhead, security, pinball problem/inefficient routing, lack of multi-homing support and mobility management. Seeking for novel paradigms, this section hopefully endows a clear sight synopsis of the diverse methods, especially those who attempt to offer an integrated alternative to all the problems and challenges of supporting QoS in moving networks.

In crowded motion environments, the bottle neck is usually occurs between the access link and the mobile network. In this setting, prioritizing traffic and assigning a least bandwidth assurance to each user is essential. The resource supervision in the mobile network is a challenging issue due to the restricted and inconstant wireless link bandwidth. Noor and Edwards [19] have proposed an active QoS provisioning structure for user-specific traffic differentiation. Resources were provisioned between the mobile router and mobile network nodes. Also, traffic was prioritized according to the user classes. The user class technique supplied guaranteed bandwidth for chosen traffic classes, although the usage of link bandwidth was worse. However, the concept of route optimization is barely delivered in the model.

Another attempt by Yu and Tao [20] to suggest an enhanced FHMIPv6-based approach that primarily carries out joint recognition function for a Mobile Node (MN) and Mobile Anchor Point (MAP). The proposed scheme computes the Standardized Edit Distance to examine the gesture provisional and guesstimate the gesture pattern of the mobile node. According to the judgment ofthe motion pattern, the MAP can regulate how MN connects to the new access point in order to avoid reusing some prior handover information and redundant binding updates. This scheme enhances the intact network performance in terms of packet loss rate, handover latency, and the throughput, particularly when the MN with the ping-pong motion pattern. Nonetheless, in the numerical results, the authors could compare and apply other redundant binding update schemes apart such as FMIPv6 and HMIPv6.

Slimane et al. [21] proposed an autonomous infrastructure interpretation. It entails no alteration in Internet architecture while maintaining seamless connectivity to satisfy the real-time applications and QoS. In order to take mobility decisions from the mobile router (MR), NEMO Gateway (NMG) entity is introduced. NMG is in charge of managing handover and allocating traffic among the MRs in a transparent manner for the mobile network nodes within the NEMO network. The traffic of the mobile network nodes passes through the NMG and all of NEMO MRs mobile are either wired or wirelessly linked to NMG. The NMG has multiple interfaces therefore it uses the standard IEEE 802.21 with Media independent handover (MIH) services to manage the MRs. The authors introduced to adopt a proactive approach consisting in carrying out soft handover (Make Before Brake) to produce availability of several tunnels simultaneously which will be used based on the management of mobility. Therefore, traffic swipes through another MR with a prior defined tunnel in case of failure of an active link for a specified MR. This work has been regarded as a multi-homed NEMO network with numerous MRs resided in various HAs, however, the multi-tunnelling will increase the processing overhead and the packet header size. In addition to that, any failure on the NMG will impact adversely on the entire network.

An innovative two-level aggregation-based QoS architecture was introduced by Wang et al. [22] as a feasible alternative for NEMO network QoS provision. In order to handle the QoS effectively, QoS accumulation and SLS co-operation were implemented at node and network level. A signalling protocol used to exchange data either between the NEMO the home/visited network domain or note/network levels. While this research work analyses the problems experienced by moving networks when offering QoS, more detailed implementation of the signalling protocol is required.

The Host Identity Protocol HIP is used to separate a host's identity from its location on the Internet. The HIP-based micro-mobility management resolutions tailored to the NEMO situation that did not enclose all safety demands and yet suffer from safety deficiencies Thus, Smaoui et al. [23] proposed a new secure and efficient scheme for network mobility management to identify a number of potential threats in the typical HIP. The proposed scheme enhances well-built verification between network entities, certifies end-to-end concealment as well as integrity protection. It also decreases Denial of Service attacks and safeguards against Domain Name Servers.

Chen et al. [24] studied the Internet connectivity of multiple-interfaces at Mobile Router (MR) in order to use various network resources effectively and make use of different Internet Service Providers (ISPs). The MR equipped with (WLAN, CDMA, and GPRS) at egress interfaces simultaneously. An interface handover choice algorithm for seamlessly transfer of data between distinct interfaces is suggested. The findings collected show thatthe multi-interface system not only promotes big area movement across varied MR networks but also offers a seamless transfer with no loss of packets and little time to disrupt service.

Labiod and Tla is [25] proposed a new resource reservation protocol called NEMO Reservation (NEMOR) to support QoS guarantee in NEMO context. The authors used generic signalling protocol called NSIS that exploits the advantages of both IntServ and DiffServ models to provide a suitable QoS to NEMO. This work builds a virtual tunnel of reserved resources to ensure the aggregation of various flows using RSVP. In two stages the claim statement is implemented: the HA-CN and MR-HA. NEMOR improved significantly the performance in terms of service continuity and signalling load. Nevertheless, the NEMOR protocol published as an Internet draft and was not went further. Also, no performance evaluation study was conducted in order to appreciate the benefits and gauge their associated overhead.

In order to eliminate excessive binding updates for MN's CoA, HaneulKo et al. [26] have figured out a delayed location management (DLM) approach where an MN defers its binding update while waiting for a pre-described timer T expires. However, the authors have solely developed an analytical model without bearing in mind simulation for evaluation. Energy costs should be considered in the analytical evaluation.

The NEMO basic support protocol often performs the same operations, regardless of a mobile network's characteristics. Therefore, it cannot achieve optimal performance. An adaptive NEMO support scheme based on hierarchical mobile IPv6 (HMIPv6) has endorsed by Pack et al, in order to diminish tunnelling overhead [27]. The general idea is a trade-off between adaptive binding update traffic and tunnelling overhead. Depending on the session-to-mobility ratio (SMR), the suggested scheme improves binding update (BU) traffic as well as tunnelling overhead by using the adaptive binding update technique. This SMR has been contrasted with distinct binding update processes with a preordained threshold. Analytical approaches have been established in contradiction of other NEMO support protocols to assess the efficiency of the adaptive NEMO support protocol. Similarly, an ideal threshold has been obtained for adaptive binding updates. Extensive numerical findings show that the adaptive NEMO support protocol achieves a substantial increase in efficiency for low SMR and demonstrates similar efficiency when the SMR is increasing with the NEMO basic support protocol. However, reducing overhead and addressing security issues should be taken under consideration in the proposed protocol.

Zhu Tang et al. [28] have proposed a new access register protocol and an active notification scheme to support the end-host and network mobility in LISP networks. The evaluation has occurred taking mobility handover delay, mobility handover overhead and packet drop rate as performance metrics. Krishan Kumar et al.[29] have suggested a spectrum handover protocol for ideal network choice by means of Multiple Attribute Decision Making (MADM) techniques in NEMO-based cognitive radio vehicle networks. Applying MADM techniques offers a broader and optimal option with quality of service amongst the access networks. The proposed protocol has been assessed against various scenarios based on internetwork spectrum handovers such as mobile network velocity, delay sensitive, access price and data rate. However, the proposed protocol should also be considered a technique such as self-organizing NEMO-based CR vehicle networks that learn thoughtfully the coordination of any network circumstances.

Hybrid centralized-distributed mobility management architecture in the context of NEMO was proposed by Nguyen and Bonnet [30]. Distributed Mobility Management (DMM) has been developed as a brand-new solution to defeat the centralized of mobility management protocols. However, there are several drawbacks must be dealt with in the proposed scheme DMM such as high signalling cost, tunnelling management and handover latency for rapid users. The proposed solution (H-NEMO) allows the devices to obtain connectivity either from fixed or mobile locations while keeping their ongoing flows. The arithmetical findings exhibited that H-NEMO outperforms NEMO BS protocol in terms of signalling cost, packet delivery cost, handover latency and end-to-end delay. However, more performance metrics need to be well-thought-out concerning to the impact of the number of nodes and flows.

Wang et al. [31] introduced a different approach for a Bandwidth Reservation Scheme (BSR) so that a Network Mobility (NEMO) can support QoS services. The proposed BSR approach can support mobility and efficiently manage reserved bandwidth as well. The suggested BSR scheme primarily deploys the notion of HMRSVP to make passive reservations on neighbouring networks in advance. When the mobile network links the present access router with its Home Agent (HA), the active reservation is initiated, whereas the passive reservation is made from the neighbouring access router and HA. The reserved tunnel between MR and its HA used to increase the bandwidth utilization in the BSR proposed scheme. It correspondingly reduces the signal overhead induced by autonomous end-user reservations. The BSR has designed three reservation adjustment policies schemes to further improve bandwidth utilization. The mobile network has three options that may select a static, dynamic or mixture bandwidth reservation. These schemes were assessed and tested via a scientific analysis and simulation as well. The assessment shows that when the mobile network is overloaded, the probability of session dropping decreases. For a dynamic policy, the reservation usage is higher than for a static policy. However, the simulation setup scenario for BSR approach is superficially discussed.

## 4. THE PROPOSED SCHEME

In this section, a novel solution called (Diff-FH NEMO) is presented. The proposed scheme provides QoS with mobility management to end-users in roaming network. It also enhances route optimization (RO) within network mobility to enable direct path communication between any kind of MNN and CN in the Internet. An overview of the proposed scheme is briefly delivered, before explaining in detail how the proposed solution works.

### 4.1. Description of the Proposed Scheme

QoS provisioning requires the deployment of various functions such as QoS establishment, admission control, QoS negotiation and renegotiation, resource management, overcrowding control, signal protocols, routing protocols, traffic conformance control, QoS supervision at various spots in the network particularly at routes. Existing communication networks are very different from one to another in terms of topology, complexity, physical media, protocols, covered distances, management architecture and police, targeted application and so on. Therefore, the implementation and administration of QoSvary among network classes. Although there is plenty of researches that can be constructed to support QoS in NEMO using application layer or physical levels or somewhere in between, this paper focuses to consider QoS support in network layer three (L3) in the form of guaranteed service.

The proposed so-called DiffServ Fast Hierarchical Network Mobility (Diff-FH NEMO) scheme was initially introduced in [32]. This framework designs to integrate the existing QoS over IP architecture with NEMO BS protocol. The goal is to satisfy both assured QoS and communication portability requirements. Captivating in mind the advantages of the Differentiated Service (DiffServ) method, our aim is to suggest the requisite adjustment in NEMO to ensure smooth portability can be accomplished with the appropriate QoS parameters.

The topology portrays underneath in Figure 3, it based on an IPv6 network with mobility management support and DiffServ model composite within NEMO to offer privilege QoS guaranteed service. The entire network moves literally as asingle unit. The proposed topology enables nodes to clearly be bound to the network as if they had been attached to a fixed point and masks the difficulty of the mobility treatment on Mobility Anchor Point (MAP), DiffServ Mobile Router (DMR) and Home Agent (HA) as well. The HA is a fixed entity in charge of communication between CN and MNNs via DMR. The DMR stays where the moving portion in the topology. It can use whatever communication technology available (e.g. 3 G, 4 G, 5G, Wireless LAN and Wi-Fi) to guarantee the rout ability. The suggested scheme presupposed that the DMR has the capabilities of the Edge Router (ER) in the current framework, and will, therefore be allowed to enforce the policy.

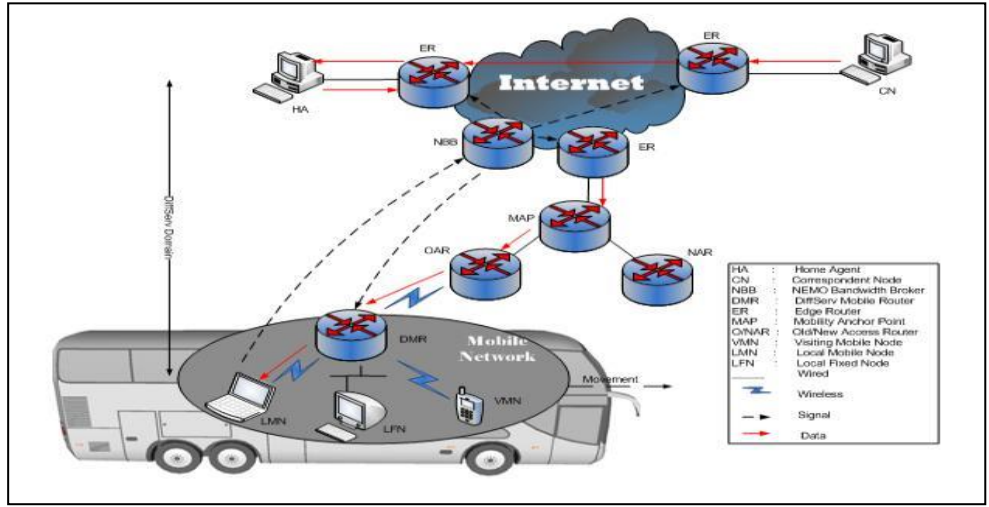

Figure 3. The suggested micro / intra motion network topology

The MAP is acting as a local HA to reduce the location update delay. The DMR collects router advertisements from Access Routers (ARs) that comprehend information regarding local MAPs when accessing a MAP domain. Next, there are couples of care of addresses (CoAs) are provided by the DMR: an on-link Local CoA (LCoA) and a Regional CoA (RCoA) in the interior of the particular MAP domain. The RCoA defines a precise Internet sphere. It is basically an address on the MAP's subnet and known as a global address. In comparison, LCoA is defined as an exceptional local address inside the domain. It is formed on the prefix posted by the AR, which adjusts at any moment the DMR updates the present AR. Indeed, by decomposing the network in several domains, a DMR doesn't need to update the HA/CN when it roams within the same domain. The NBB serves as a bandwidth broker to control and track the network capacity. Through the distribution of resources to the forwarding groups and managing the traffic volume for these categories, the proposed scheme would generate different rates of quality and resource reliability, but not complete bandwidth assurances or delays to specific flows. NEMO Bandwidth Broker is explicitly accountable for routing packets to one of the forwarding groups sponsored on the network. It should certify that individual users ' traffic obeys the SLA. This SLA extracts from the Traffic Conditioning Agreement (TCA). The SLA defines the treatment for in-/out-of-profile traffic flow behaviour, re-marking laws, and traffic profiles. The edge nodes transform the TCA into each client's traffic profile. When the packets cross the edge nodes inside the network, the resource distribution will be performed depending on the category of forwarding.

The Mobile Network Node (MNN) determines the specifications by using SLA to query services. The contrary, the NBB operator has to undertake an admission regulation role by accepting or rejecting a demand for bandwidth. It manages a list of conditions according to which reservations are made and then assigns a DiffServ Code Point (DSCP) to those services. Bandwidth sharing deals must first take place between the MNN and the NBB. Eventually, if the NBB approves the service demand of the QoS, the edge routers and the DMR will be designed to help leverage existing resources (i.e. network load control) as shown in Figure 4.

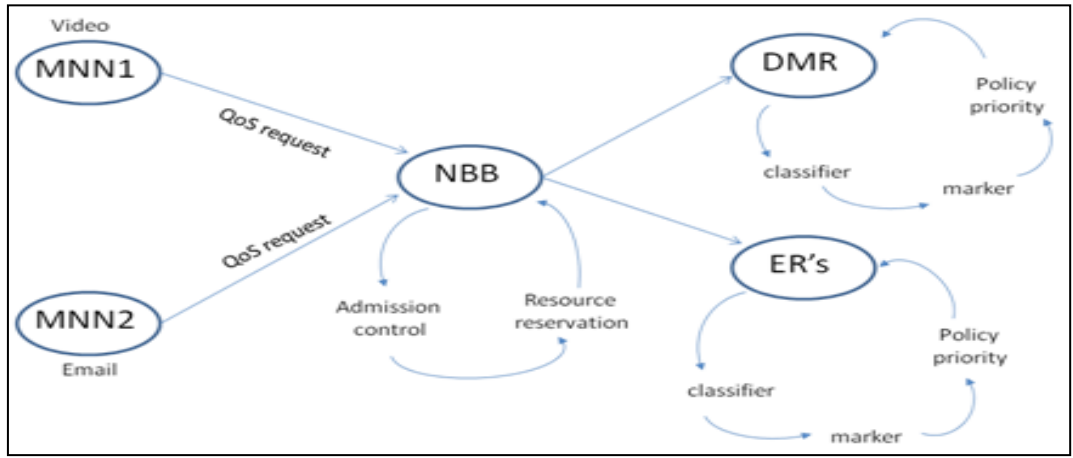

Figure 4. Traffic classification and condition in the proposed scheme

## 4.2. The proposed scheme operations

The proposed scheme has two generic modes of operation, predictive and reactive mode. Both modes operate either when the DMR moves with in the same MAP domain (i.e. Micro/Intra Mobility) or moves into a new MAP domain (i.e. Macro/Intra Mobility).

### 4.2.1. Predictive mode transactions of the proposed scheme (Diff-FH NEMO) in Micro/Intra mobility swap

A Comprehensive explanation of the operating actions and the series of signals utilized in the proposed scheme are given below and exemplified in Figure 5:

As far as we know, the data on the lower layer (L2 trigger) is either used to predict or to speed up events in the handover of network layers. DMR can only check to identify the Access Point (AP) with a stronger signal to move to, without realizing if the AP is connected to a different AR. Only the link layer knows address and AP names. Thus, the L2 handover expectation leads the DMR to realize its movements to NAR. Also, the DMR will be able to obtain a potential new Address Care (CoA) while still linked to the ancient AR. It has to check the validity of this address on the new link (i.e. Duplicated Address Detection). Yet, this can't be completed until it roams to the new AR.

- ✓ The DMR dispatches router solicitation for proxy (RtSolPr) signal to old AR to demand current AR data that is known as a new on-link care-of address (LCoA). It must comprise data about the NAR network prefix, NAR network identifier, and link-layer address.
- ✓ Upon receipt of RtSolPr, the OAR responds with the extended Diff-FH NEMO PrRtAdv signal that includes the NAR network prefix, QoS awareness data and Internet Protocol address.
- ✓ By the means of prefix information, the DMR will configure new CoA (i.e. NLCoA) using stateless auto-configuration or stateful configuration.
- ✓ The DMR directs high-priority Fast Binding Update (FBU) signals to MAP to connect former PLCoA to current NLCoA.
- ✓ The MAP must begin the handover phase after obtaining the FBU notification from DMR and deliver a Handover Initiate (HI) signal to the NAR. The HI signal contains an application for confirmation of pre-configured NLCoA and a bi-directional tunnel to avoid routing failure at handover.

- ✓ In answer to the HI signal, NAR tests NLCoA's availability via DAD (Duplicated Address Detection) and customs HACK (Handover Acknowledgement) to create abi-directional tunnel to MAP.
- ✓ The MAP dispatches signals of Fast Binding ACK (FBACK) over PLCoA and NLCoA to the DMR. This FBAck signal keeps the validation record status of the pre-configured latest CoA and tunnel initiative to DMR.
- ✓ The MAP would connect the ancient with new CoA and channels any packets destined to PLCoA towards NLCoA over NAR's link. Such forwarded data packets, labelled with a Differentiated Service Code Point (DSCP) tag, are buffered by the NAR into IPv6 packet headers up until the DMR connects to the NAR link.
- ✓ By conveying Router Solicitation (RS) signal to NAR with the Fast Neighbour Advertising (FNA), the DMR declares that it will be on the new link. The buffered packets are then sent to the DMR by NAR. In contrast, the DMR decapsulates and transmits DSCP data packets to the planned MNN.
- ✓ The DMR derives the RCoA from the MAP site and the existing AR contents aLCoA. Subsequently, if DMR desires to let go ofthe link, it sends to MAP a high-priority Local Binding Update (LBU) comprising both RCoA and LCoA to establisha bond between both addresses. The LBU also has a prefix for the mobile network (MNP). When MAP receives LBU from DMN with the new LCoA (NLCoA), it cuts packets short from NAR and then clears the tunnel that has been created.
- ✓ The MAP must submit local binding ACK (LBAck) to the DMR in reaction to LBU.
- ✓ The MAP keeps the bond in the Binding Cache (BC) and sends the LBU to the DMR's HA and CN to remind them about the current RCoA. It is also earned in exchange (LBAck). Furthermore, as long as DMR travels in the same MAP domain by separate ARs, it does not have to remind the HA or CN of their motion (i.e. new LCoA), as the RCoA remains unchanged and only the LCoA has been updated.

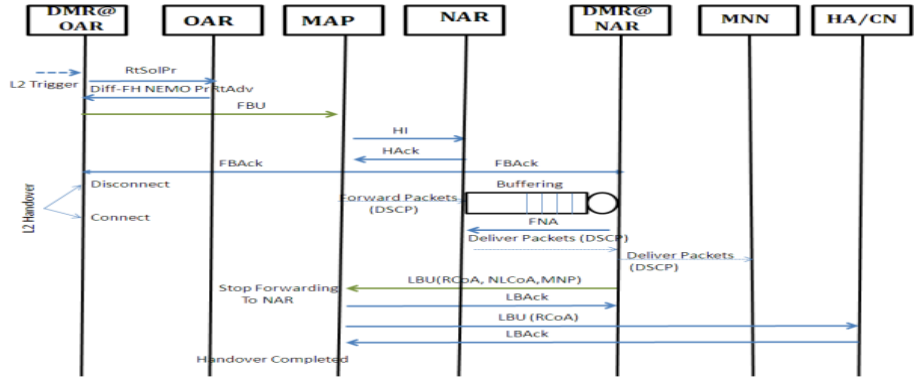

Figure 5. The signals order of the proposed scheme in the predictive micro/intra regime

**4.2.2. Reactive mode transactions of the proposed scheme (Diff-FH NEMO) in Micro/Intra mobility swap**

The reactive mode is only a complement to the prior predictive mode. It is commonly resolving different fault / inaccurate circumstances. The reactive mode process is displayed in Figure 6. The unexpected failure of the connection leads the DMR to lose its synchronization to the AR and forcing the link layer execution to move to new AR before the expected time. Therefore, due to the rapid travel of the DMR from the preceding link to the current link, there are two forms of fault / inaccurate circumstances:

International Journal of Computer Networks & Communications (IJCNC) Vol.12, No.2, March 2020

a) FBU did not forward: the DMR expected its motion in this situation but failed to send FBU demanding that the MAP onward the packets to the new link.

b) FBack did not receive: the DMR sent the FBU in this case, but received no recognition, as after the FBU was sent, the DMR port the connection.

To resolve the ex-instance, it can be done by sending FBU with high priority to the MAP after the DMR assigns to the new AR. As far as the HI message has previously been received by the NAR, MAP is going to establish a tunnel. The FBU is therefore forwarded through the tunnel and the packets are redirected to the new site of the DMR.

The DMR is not convinced that the FBU has been effectively dealt with by the MAP, as the FBack is the only recipient sign of FBU. Hence, to fix the above inaccurate occasion:

- ✓ As long as the DMR binds to NAR, it will onward FBU per high priority compressed in the FNA signal toward NAR.
- ✓ If NAR senses that NCoA is engaged in the processing of the FNA (i.e., address collision), it will remove the internal FBU packet and dispatch a Router Advertisement (RA) message with the Neighbour Advertisement Acknowledgement (NAACK) option where NAR might embrace an alternative IP address for the DMR to be used.
- ✓ Else, FBU is forwarded to MAP by the NAR which answers with FBack.
- ✓ Just now, via the NAR connection, MAP will begin tunnelling any packets that are forwarded to PCoA to NCoA. NAR can formerly supply those packets to the DMR.

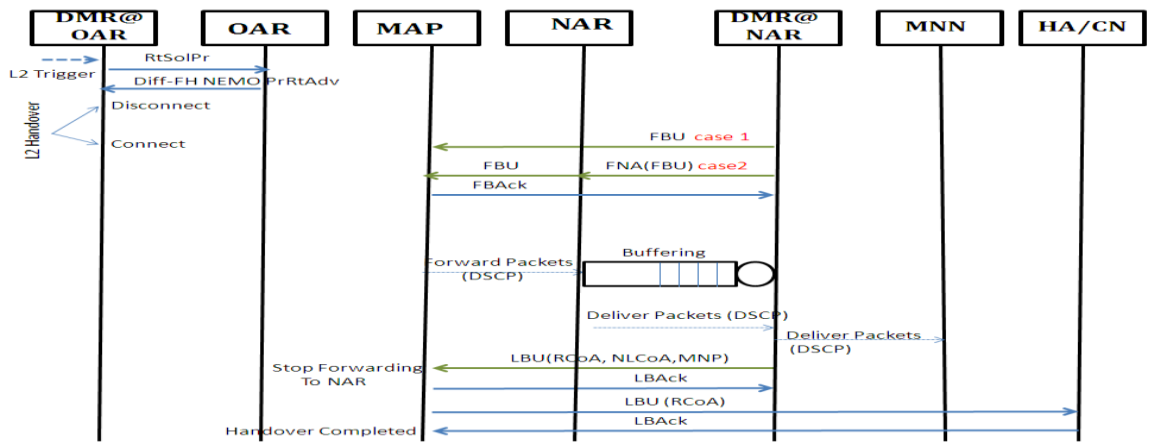

Figure 6. The signals order of the proposed scheme in the reactive micro/intra regime

### 4.2.3. Operation of proposed scheme (Diff-FH NEMO) in predictive mode of Macro/Inter mobility handover

In the prediction macro/inter mode for the proposed scheme, the DMR moves from outside the old MAP domain towards a new MAP domain as depicted in Figure 7. The HA and the CN have to be informed of the change in this situation unlike the previous micro/inter mobility handover. The proposed scheme exploited the use of route optimization support for Mobile IPv6 available in the Correspondent Nodes to provide route optimization for mobile networks. Therefore, when DMR sends a BU to the CN that is situated within the site, the CN would be able to send data packets marked with DSCP into IPv6 packet header directly to the DMN and then to intended MNN without MAP or HA intervention.

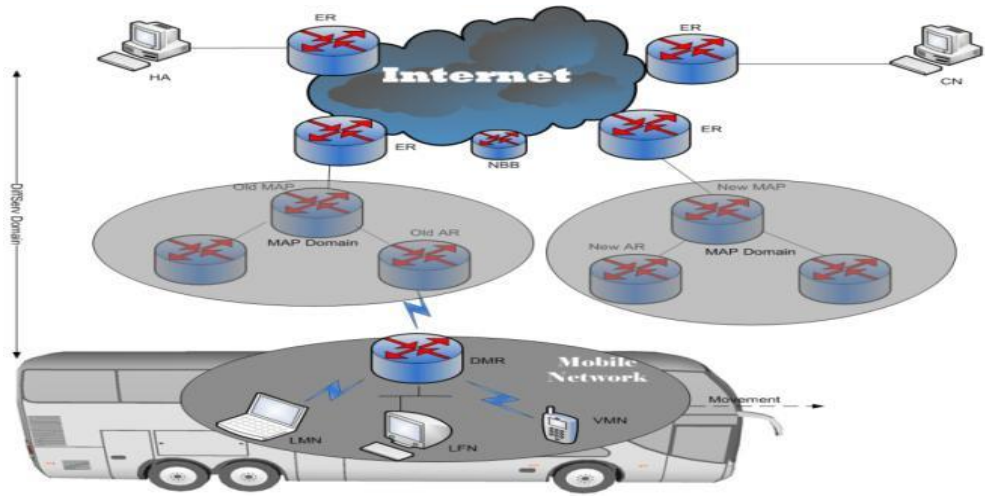

Figure 7. The proposed network topology in macro/inter movement

The operation procedures in the prediction macro/inter mode, are mostly identical to micro/inter except that the bi-directional tunnel is established between old MAP and NAR located in the new MAP domain. The sequence of messages/ control flow in the prediction macro/inter mode are illustrated in Figure 8. The detailed description of operating procedures is summarized below:

- ✓ Based on Layer 2 handover anticipation, the DMR sends proxy router solicitation (RtSolPr) message to OAR for neighbourhood prefix discovery (i.e. in order to obtain the network information immediately).
- ✓ The OAR directs the extended Diff-FH NEMO PrRtAdv message to the DMR in response to the RtSolPr message, which includes data on the new DMR's LCoA to be used in the NAR district and information on QoS awareness. So, the DMR would anticipate the occurrence of the macro mobility handover and the necessary traffic profile.
- ✓ Using the information involved in the received Diff-FH NEMO PrRtAdv message, the DMR generates both a New LCoA depended on the prefix of the NAR and a New RCoA depended on the prefix of the New MAP.
- ✓ Then, DMR sends a Fast Binding Update (FBU) message with high priority to the old MAP in order to bind the previous PLCoA with new NLCoA.
- ✓ Once the old MAP received FBU message from the DMR, it will send a HI message to the new AR that the DMR expects to move to. The NAR must perform a Duplicate Address Detection (DAD) process for the NLCoA requested by the DMR. Concurrently, the NAR also sends a HI message to the new MAP in order to request the DAD test for the new RCoA asked by the DMR. In response to the HI message, a HAck message should be sent back to the NAR and old MAP at the same time.
- ✓ Next, through PLCoA and NLCoA, the ancient MAP sends Fast Binding ACK (FBACK) signals to the DMR. At this very moment, the layer 2 handovers would take place. As a result, the old MAP establishesa bi-directional tunnel to NAR with all the data packets that have been marked with a Differentiated Service Code Point (DSCP) value into IPv6 packet headers.
- ✓ The MAP commences forwarding the data packets toward the NAR by using the established tunnel.The data packet can be kept in the NAR for some time up until the DMR notifies NAR of its attendance on the new link by transferring Router Solicitation (RS) signal along with the Fast Neighbour Advertisement (FNA) signal. Consequently, NAR delivers the buffered packets to the DMR. Therefore, packet loss can be reduced in macro mobility handover. Moreover, the DMR decapsulates the tunnel and forwards DSCP data packets to intended MNN based on MNP.

✓ The DMR should perform the local registration by using Local Binding Update (LBU) with high priority and Local Binding ACK (LBA) messages with the new MAP. The LBU includes new LCoA, new RCoA at the new MAP and MNP.

✓ When the new MAP receives the LBU, it will stop the packet forwarding to NAR and then clear the tunnel established from the old MAP for fast handover. The MAP supplies the bindings in the Binding Cache (BC) as well as forwarding the LBU to the HA and CN of the DMR to inform them of the current RCoA. Then, it will receive (LBAck) simultaneously.

✓ Finally, after completion of the local registration in Macro/Inter mobility handover, the DMR must send Binding Update (BU) message with high priority to the HA in order to inform about its movement toward new MAP with New RCoA. So, the BU will bind the DMR home address with its NLCoA (which is located in the new MAP with new RCoA). In order to enable route optimization in the proposed scheme, BU procedure must also be performed to CN. Nevertheless, return routability (RR) procedure test must be performed before executing a binding update process at CN in order to assure that BU message is authentic and does not originate from a third party/ malicious DMR. The procedure of the Return Rout ability process is briefly expounded as following points: By tunnelling the signal over the HA, the DMR leads a Home Test Init (HoTI) signal indirectly to the CN. The DMR then sends a signal from the Care-of Test Init (CoTI) straight to the CN. After that, in reaction to the HoTI signal (sent indirectly to the DMR over the HA), the CN dispatches a Home Test (HoT) signal carrying (Home Keygen). Likewise, the CN dispatches a Care-of Test (CoT) signal carrying (Care-of Keygen) in retort to the CoTIsignal (sent directly to the DMR). Afterward, the correspondent node sendsthe Network Prefix Test (NPT) message with (Prefix Keygen) indirectly to the DMR via the home agent [33]. So, now the CN can only be capable to send the packets optimally direct to the DMR instead of going through the HA or MAP (i.e. the CN would be confident to verify that the care-of address and home address of the DMR are collocated as well as it will make sure that the DMR indeed owns the network prefixes it claimed to own).

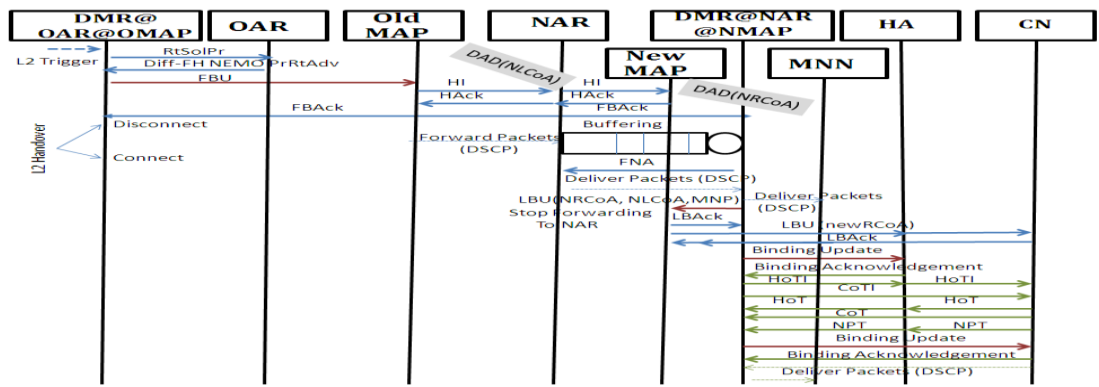

Figure 8. The messages sequence in the prediction macro/inter mode for the proposed scheme

### 4.2.4. Operation of proposed scheme (Diff-FH NEMO) in reactive mode of Macro/Inter mobility handover

The reactive mode of macro/ inter mobility handover for the proposed scheme is likewise to the reactive mode of micro/intra mobility handover as in section 2. It is also needed to solve the various erroneous cases and maintain improving the fast handover operation. The procedure of the reactive mode of macro mobility handover is shown as Figure 9.

If the DMR does not manage to send the FBU message or receive FBAck message prior to the layer 2 handover, it will convert the fast handover operation into the reactive mode of macro mobility handover.

As soon as the DMR joins to the new AR, it is able to launch FBU to the MAP in case of FBU gets lost (i.e. first erroneous situation). Since the HI message has already been obtained by the NAR, MAP is going to establish a tunnel. Therefore, the FBU is forwarded through the tunnel and the packets are redirected to the new place of the DMR.

The contrary, if FBack gets lost because of the fast motion of the DMR from the former link to the new one, on this occasion the DMR will forward FBU encapsulated in the FNA message to NAR (i.e. in the occasion of the second erroneous situation).

- ✓ If NAR senses that NCoA is engaged in the processing of the FNA (i.e., address collision), it will remove the internal FBU packet and dispatch a Router Advertisement (RA) message with the Neighbour Advertisement Acknowledgement (NAACK) option where NAR might embrace an alternative IP address for the DMR to be used.
- ✓ Else, FBU is transmitted by the NAR to MAP replying with FBack.
- ✓ Finally, the sequence of procedures will follow the same steps of predictive mode of Macro/Inter mobility handover. The MAP can start tunnelling any packets addressed to PCoA towards NCoA through NAR's link. The NAR can then supply these packets to the DMR. The DMR should perform both the local binding operation by using LBU and LBA messages and the location binding operation by using BU and BA messages with its HA as well as CN.

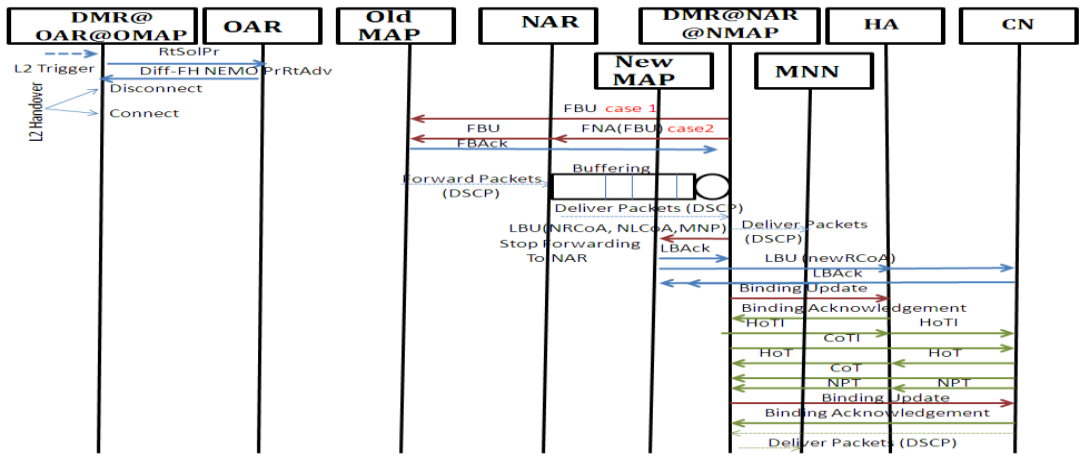

Figure 9. The messages sequence in the reactive macro/inter mode for the proposed scheme

**4.2.5. Operation of the proposed scheme devoid of mobility management (Diff NEMO)**

As we know that, in NEMO Basic Support Protocol (NEMO BSP) the packets to and from the mobile network travel through a bidirectional tunnel [34] between a Home Agent and a Mobile Router (HA-MR). Encapsulation and tunnelling packets via the HA outcome in amplified header overhead and inefficient route issues. This causes the deterioration in the efficiency of real-time applications. The problem evolves even further with increasing nested level which causes

37

multiple encapsulation packets to travel through multiple HAs. Incompetent routing is widely recognized as the "pinball routing issue" in nested NEMO networks. Hence, the proposed scheme optimizes location update traffic and tunnelling overhead by employing the standard MIPv6 mechanisms to allow upfront communication on the Internet between any type of MNN and CN. It intends to provide QoS with seamless mobility during communications among the users of the mobile network node. In theory, informing the CN with CoA and HoA of the MNN will involve major changes in the CN structure. Thus, the idea of the proposed scheme is to enable a DMR to behave as a proxy for any MNN, by trading the packets needed by MIPv6 route optimization. The mechanism of the proposed scheme (Diff NEMO) involves three phases: address configuration, binding update procedure, and packet delivery procedure.

### 4.2.5.1. Configuration of Address

The DMR intuitively moves from Old AR to a New AR when it performs handover procedure as shown in Figure 10. It is used to maintain two addresses: permanent address which is used when the DMR in its home network and temporary address which is used when the DMR in the visited network. The stable address is named Home address (HoA) whereas the momentary address is referred to as care of address (CoA), which is utilized to deliver data on the present place of the DMR. The control flows for the proposed scheme is illustrated in Figure 11. At the first three steps, the MNN sends data packets to CN via the proxy DMR. The DMR marks a Differentiated Service Code Point (DSCP) value into IPv6 packet headers and forward packets to the next hop. The only way for the DMR to know that it has been moved from subnets to another (or from OAR to NAR) is by performing movement detection (MD). Router Solicitation / Advertisement (RS / RA) signals are swapped to discover anew access router (NAR). If the DMR recognizes that the link prefix is altered, the flow will be determined by DMR. The stateful and stateless method configuration are used to give the DMR a new CoA as it instantly roamed to a new sub-network. Furthermore, the Duplicate Address Detection (DAD) operation is conducted by swapping Neighbour Solicitation / Advertisement (NS / NA) signals to guarantee that a configured CoA is only one of its kind in the existing link.

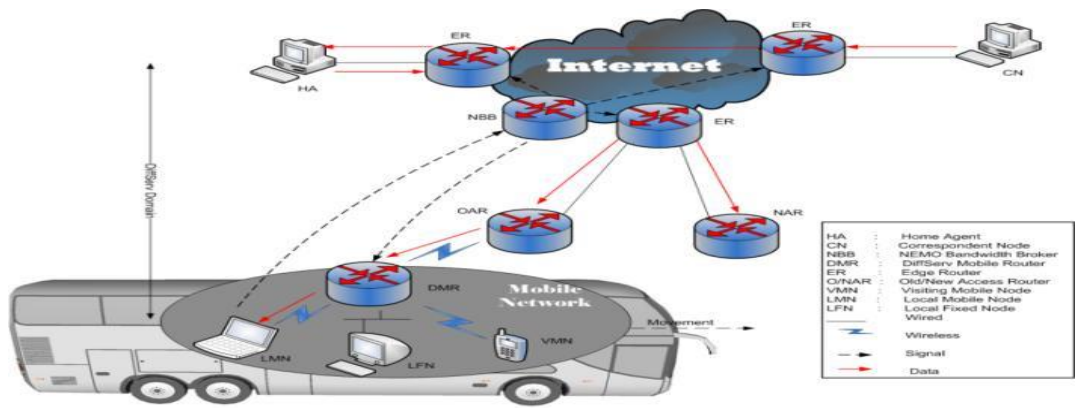

Figure 10. The proposed network topology in the proposed scheme (Diff NEMO)

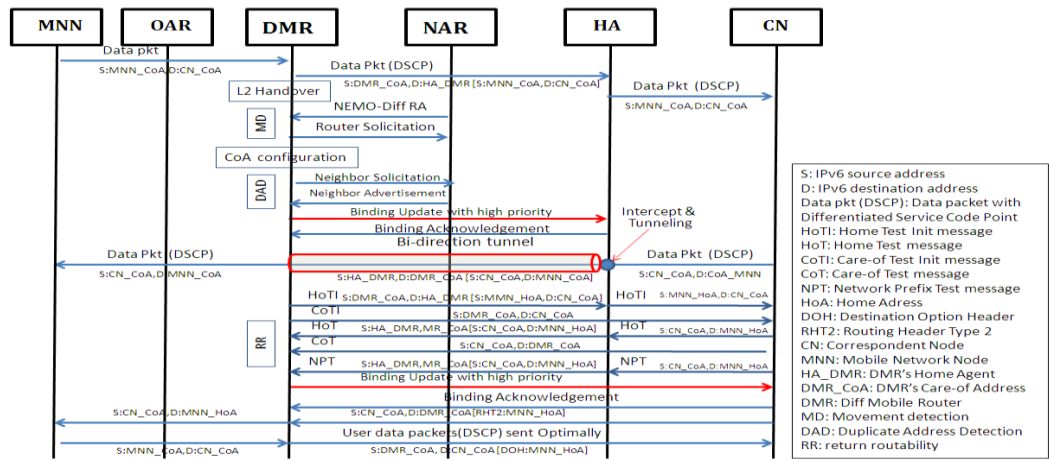

Figure 11. The sequence messages in the proposed scheme (Diff NEMO)

**4.2.5.2. Binding Update Process**

Upon the creation of a CoA, the DMR records the CoA with the Home Agent (HA) by submitting a Binding Update (BU) signal as reminding of its motions. Likewise, the latest (NEMO-Diff RA) message will be used by the DMR to advance BU with the highest priority. Furthermore, the Mobile Network Prefix (MNP) should be updated by the DMR as it frequently envoys multiple address prefixes utilized by the MNNs within its network. Subsequently, such data must be placed in the HA's binding cache to pass on packets from CN to the DMR's home address. This data is also deposited by the CN in its binding cache. The CN and HA generally uphold the binding cache. Every single entry in the binding cache comprises the home address of the DMR, the CoA, the MNP and the lifecycle representing the entry's validity. Thenceforth, the HA sends Binding Acknowledgment (BAck) signal reporting that conveying to the DMR is arranged. It also produces a binding cache entry mapping the HoA and DMR prefixes to the DMR CoA.

In compliance with NEMO terms [35], when the mobile network connection switches because of mobility, DMR conducts a handover process to preserve visibility of the movement of mobile networks from MNNs. Thus, the DMR should carry out the Mobile IPv6 Return Routability (RR) procedure as an alternative to all MNNs. To empower route optimization for the proposed scheme, BU procedure has to be done also to all active CNs. Nevertheless, according to guarantee that the BU signal is genuine and does not come from a malevolent DMR, the RR procedure should occur conclusively ahead of sending a BU signal to the intended CN. Namely, the RR technique is considered to authenticate that the DMR is accessible at home and foreign address. To avoid spoofing of binding updates, the home address must be checked (i.e. the BU is rendered by DMR but not by other parties). The CoA should be checked to safeguard in contradiction of denial-of-service outbreaks in which the corresponding node is trapped into flooding of a misguided foreign addresses with packets. In short, the method for return routability is constructed on the home address test (i.e. Home Test Init (HoTI) and Home Test (HoT) signals trading) and care-of address test (i.e. Care-of Test (CoT) and Care-of Test Init (CoTI) signals trading). The RR process is shortly exemplified as following themes as shown in Fig 11:

- ✓ By tunnelling a signal viaa domestic agent, the DMR transmits a Home Test Init (HoTI) signal incidentally to the CN. Also, it will send along with the network prefixes it possesses.
- ✓ The DMR directs a signal of Care-of Test Init (CoTI) to the CN.

- ✓ The correspondent node sends indirectly to the DMR via the home agent both of a Home Test (HoT) message (which is involved Home Keygen) in response to the HoTI message and Network Prefix Test (NPT) message (which is involved Prefix Keygen).
- ✓ In reaction to the CoTI signal (that sent straight to the DMR), the CN dispatches a Care-of-Test (CoT) signal (which encompasses Care-of Keygen).

Eventually, the DMR can securely register the binding update message with the CN (forwarded with high priority) and receive the Binding Acknowledge (BA) in return.

### 4.2.5.3. Packet Delivery Process

After the registration with the HA has been completed, the communication steps proceed by receiving tunnelled DSCP data packets at the DMR, which have been originated by Correspondent Node (CN).This only occurs once either a CN has no DMR binding (correspondent recording is ongoing) or Mobile IPv6 is not borne. As soon as the DMR records the binding update signal with the CN (i.e. including HoA of the DMR, CoA of the DMR and Mobile Network Prefix choices) and obtains the BA as a response, in the meantime the CN can ideally coordinate packets straightforward to the recent CoA of DMR avoiding the HA. It infuses the course to its steering table so that packets ordained for addresses in the Mobile Network Prefix will be sent via two-ways tunnel. This is done to prevent routing traffic through the home network, which could likely lead to a bottleneck. In the proposed scheme the DMR is held responsible for illegal traffic sent from/to its mobile network nodes. The Correspondent Node (CN) could be a stationary node or mobile node with a CoA (i.e. CN_CoA). The CN will include also DSCP data packets with Type 2 Routing header extension that contains the home address of the MNN (MNN_HoA). Therefore, the minute the DMR gets the packet, it forms Type 2 Routing header by evacuate it and then consistently trades the endpoint address of the packet from the DMR's care-of address to the home address (HoA) of the MNN. From the application layer perspective, the data is addressed from the correspondent node address to the home address of the MNN. In the same way the DMR sends the packets from its care-of address to CN and it will include Destination Options extension header which contains Home Address option that evolved home address of the MNN. If the CN gets the packet, it reasonably substitutes the packet's origin IPv6 address (the DMR's and MNN's care-of-address) with the home address deposited in the Home Address option. Again, from the application layer perspective, the data is addressed from the home address of the MNN to the correspondent node address. So, finally the users of the applications will think that, they are still communicating with CN from the MNN's home network. This provides uninterruptable connection to the users of the MNN with QoS assurances. To generate numerous packet classes in the proposed scheme, inward and outward packets will be marked in a different way with certain code.

As long as, the DMR is closer to MNN than the Home Agent, the route between MNN and the CN can be considered as an optimal route. The proposed scheme exclusively relies on the existing IPSec to secure all of signalling messages and tunnels (using IPSec Authentication Header (AH) and Encapsulation Security Payload (ESP) mechanisms).

## 5. PERFORMANCE EVALUATION AND RESULTS

In order to evaluate Quality of Service (QoS) within networks mobility environment, Network Simulator (NS-2) tool version 2.28 has been used [36]. The NS-2 [37] has been successfully installed under Cygwin and configured with NEMO BS patch which is an extension module from MobiWan [38] built by MOTOROLA Labs Paris in collaboration with INRIA PLANETE team as well as patch named FHMIP [39].The simulation space size has been set to 1600 m by 1600 m and the CBR data-flows are transmitted continually. A number of mobile network nodes is distributed through the network area which is connected to the DiffServ Mobile Router (DMR). Also, the Correspondent Node (CN) is connected to the wired Edge Router (ER) with an address (0.0.0). All nodes have a hierarchical address as indicated in Figure12. The address hierarchy is categorized to three divisions: domain, site and node. A domain usually is made up of several BSs, and each one of them is in charge of a site, all nodes on the site subsequently are treated alike then it will treat all nodes on the site as equal. The address hierarchy inside a site is divided to: one part is distributed for the mobile networks and another part is preserved by the BS for the dynamically attached DMR or MNNs to configure their CoAs. Mobile network possesses MNP which is also divided to one part for DMR and LFNs, and the remaining part preserved by BS that will be attaching other mobile networks and VMNs to this mobile network. Figure 12 presents the initial topology of the simulated network.

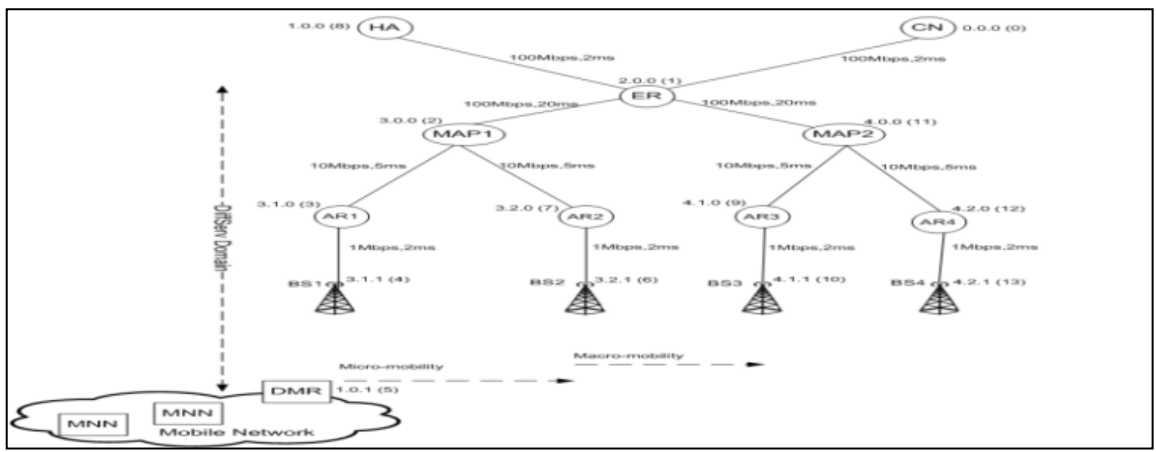

Figure 12. Topology of simulation setup

The time for simulation initiates at zero seconds and ends at 200 seconds. When the simulation gets starting the DMR is close to Home agent (HA) and launches a UDP session with CN. A "UDP" agent is attached to the Correspondent Node. The connection is established from the source "UDP" agent to the "null" agent attached to the Mobile Network Node (i.e. LFN). A "null" agent frees the packets received. A Constant Bit Rate "CBR" activity generator is joined on the upmost of "UDP" operator at the correspondent node. The "CBR" is configured to generate 1K Byte packets at the rate of 100Kbps. The "CBR" is set to start at 20 seconds and stop at 200 seconds. The CN and HA are coupled to ER with a link delay of 2 ms and the link transmission rate is 100 Mbps. The link bandwidth between ER and the MAPs is 100 Mbps link with a 20 ms link delay. Access routers (AR)s are connected to MAPs with 5ms link delay over 10 Mbps links. The ARs are further connected to the Base Station (BS) with a 2ms link delay over 1 Mbps links. The wireless technology utilized in the entire network between the base stations and the DMR is a basic 802.11. The IEEE 802.11 radius (transmission range) is 50 meters. After twenty-one seconds the DMR moves within the range of the BS1 that is belonged to AR1. Later on, the DMR signs out from the previous base station range toward AR2 with a rate of 1 meter/second (approximate human walking speed) at thirty seconds from the simulation start time. Then, the DMR carries on moving in linear motion toward AR3 and AR4 respectively where macro mobility would take place. In the simulation topology, it is assumed that the allocating and controlling of the resources have been taken under control (i.e. the MNN gets the required bandwidth in all mobile network domains). Consequently, this eradicates the need of NBB for sake of simplicity in simulation setup (since the proposed scheme scenario does not consider multi-homed or nested NEMO yet). Traffic supervision procedures used in the mock-up are, RED for buffer supervision, token bucket for entrance governor, and priority queue as scheduler in the DiffServ domain. The MNN is the destination, while the source CN is considered to be fixed node to avoid unnecessary drawbacks of simultaneous mobility problem that might happen. In case of the proposed scheme devoid of mobility management, we block out in the simulation script all of MAP_MODE, FAST_HANDOVER, MAP_FAST_HANDOVER and ATTACH-MAPAGENT. Table 1 briefly shows the simulation parameters. To study the performance of the proposed technique, we have considered the following performance metrics (in order to meet with network mobility support goals and requirements) [40]:

- **Packet Loss:** It could be interpreted as the sum of packets dropped, lost or corrupted (never reach to the intended destination) during the handover.

- **Handover latency:** It can be referred to as the passed time between the last and first packets sent from the old access point and arrived to the new access point. In another word, it is the amount of time that it takes for the mobile router to be granted access into a new network.

Table1. Simulation parameters

| Packet size | 1000bytes |
|---|---|
| IEEE 802.11 radius (transmission range) | 50 meters |
| Simulation time | 200 s |
| Traffic generator | CBR |
| MAC | Simple IEEE 802.11 |
| Admission control | Token Bucket |
| Buffer management | RED |
| Scheduler | Priority Queue(PRI) |
| Routing protocol | NOAH |
| Simulation area | 1600 x 1600 |
| Transport layer agent | UDP |

## 5.1. Packet Loss Result

The graph in Figure 13 presents a comparison of packet loss between the proposed scheme (Diff-FH NEMO), (Diff NEMO) and NEMO BS protocol. The arrival form of packets relies on various aspects such as characteristics ofthe application, behaviour of network queuing, and so on. When the mobile router speed was varied (from 15 km/h to 90 km/h) in the consecutive mobility environment such as train or bus, the packets loss increases noticeably. This result from the fact the Home Agent (HA) doesn't preserve the excessive packets that have been sent from the CN in the time of congestion as per the standard NEMO BS. The overcrowding in this situation is achieved by adjusting the network load to more than 1 Mbps in order to obtain high link congestion. Therefore, the packets that have been intercepted by HA to be sent to the intended the mobile router will be dropped until the HA will be updated with the new CoA of the mobile router and processed the BU. To minimize the packet loss the movement detection and configuration of a new CoA for the mobile router must be executed as quickly as possible. On contrary, the arriving packets from the CN via the MAP in the proposed scheme (Diff-FH NEMO) will be buffered by the new AR in advance, even before the DMR is able to establish its link there. So, the proposed scheme (Diff-FH NEMO) uses the HI message the sign for buffering at the NAR. Thus, this traffic buffering is managed legitimately to minimize the packets loss and smooth handover compared to other schemes, which is adequate to real-time application the required a certain QoS restrictions. Unlike the standard NEMO BS, traffic in the proposed scheme (Diff NEMO) is divided into small number of groups called forwarding classes. The forwarding class, that a packet belongs to is encoded in a field in IPv6 packet header. Each forwarding class embodies a limited accelerating deal of drop bandwidth and distribution precedence, which fairly improves and minimizes the overall packet loss in time of congestion for real-time applications. In short, it can be observed that the proposed scheme (Diff-FH NEMO) has relatively decreased the packets loss lesser than (Diff NEMO) and the standard NEMO BS protocol.

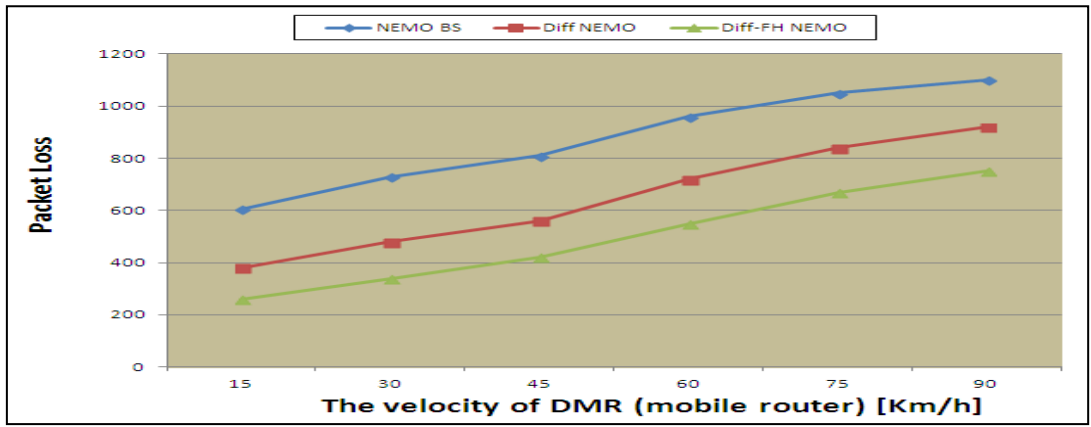

Figure 13. The packet loss versus speed

## 5.2. The Packet Forwarding Rate Result

Figure 14 shows the packet forwarding rate when the speed of the mobile router was changed (from 15 km/h to 90 km/h) in the case of the proposed scheme (Diff-FH NEMO), (Diff NEMO) and NEMO BS protocol. It can be deduced from the graph, the packet forwarding rate abruptly decreases to 48.43%, 45.27%, and 38.03% respectively, when the mobile router speed was 60 Km/h. Furthermore, the degradation of the bandwidth could become worse and reach to 25.32% in the velocity of 90 Km/h. The proposed scheme (Diff-FH NEMO) has shown a higher packet forwarding rate compared to (Diff NEMO) and the standard NEMO BS protocol.

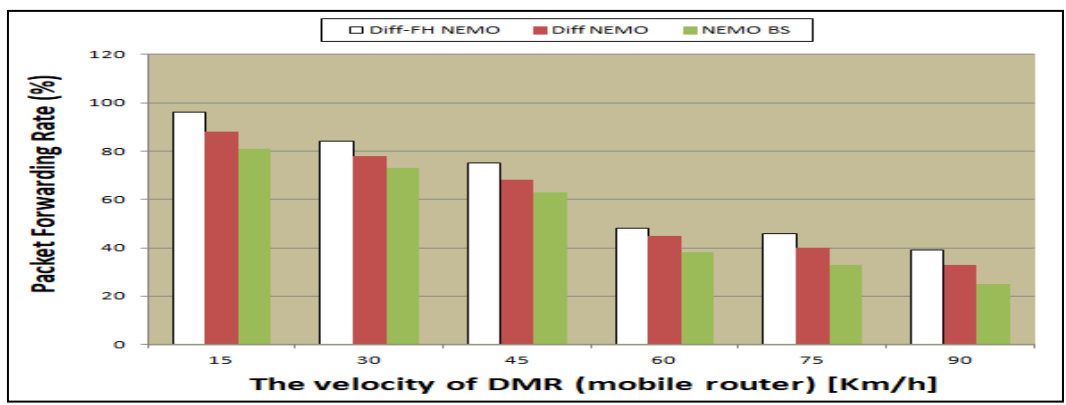

Figure 14. The packet forwarding rate versus speed

## 5.3. The Handover Latency Result

As in Figure 15, the graph studies the handover latency when the mobile router speed was adjusted (from 15 km/h to 90 km/h) in case of the proposed scheme (Diff-FH NEMO), (Diff NEMO) and NEMO BS protocol. It can be noticed that when the velocity of the mobile router that resides in mobile networks increases the handover latency will comparatively increase. The standard NEMO BS protocol has the worst handover performance compared to other schemes. The reason behind that implies in all the data packets that meant to the mobile network node or CN has to go through the Home Agent (HA). Thus, launching two-way tunnel among the home agent and the respective mobile router permits packets to flow in both directions, although the mobile router is linked to a visited link. However, it results in increased length of packet route and increased the handover latency in most cases. The sub-optimal path apparently causes various inefficiencies associated with packet delivery. Namely, handover latency, processing delay (at the points of encapsulation and decapsulation), packet overhead (that amplifies packet size due to the addition of an outer header), and bottleneck links that lead to traffic congestion and susceptibility to link failure. In this sense, all mobile network nodes (MNNs) communications can be ultimately disrupted. On the other hand, the proposed scheme (Diff-FH NEMO) reduces the handover latency that caused from home network registration. Namely, it reduces the delay associated with Binding Update (BU). As long as, the distance between the HA/CN and DMR is fairly long, the essential time for the binding update in the subsistence of the Mobility Anchor Point (MAP) will be significantly shrunk by local BU. At same time, the proposed scheme (Diff-FH NEMO) diminishes the latency associated with motion detection and verification/configuration of the CoA in order to support seamless network access services to the DMR. Furthermore, the handover delay will be reduced even more in case when the DMR has to update HA or CN to perform macro mobility because the packet will be sent through optimal route. In other proposed scheme (Diff NEMO), the data packet will be sent directly from the CN to the DMR and then to intended MNN without interference of HA or the MAP. Moreover, it exploits two kinds of forwarding classes mechanisms (service models) such as Assured Forwarding (AF) PHB and Expedited Forwarding (EF) PHB to diminish the handover latency for real-time applications like video telephony and virtual conferencing. So, the proposed scheme (Diff NEMO) mostly improves the handover delay between the MNN and CN, with an extra benefit of reducing the traffic of the Home Network. As a result, the proposed scheme (Diff-FH NEMO) reduces handover latency less than (Diff NEMO) and the standard NEMO BS protocol, respectively.

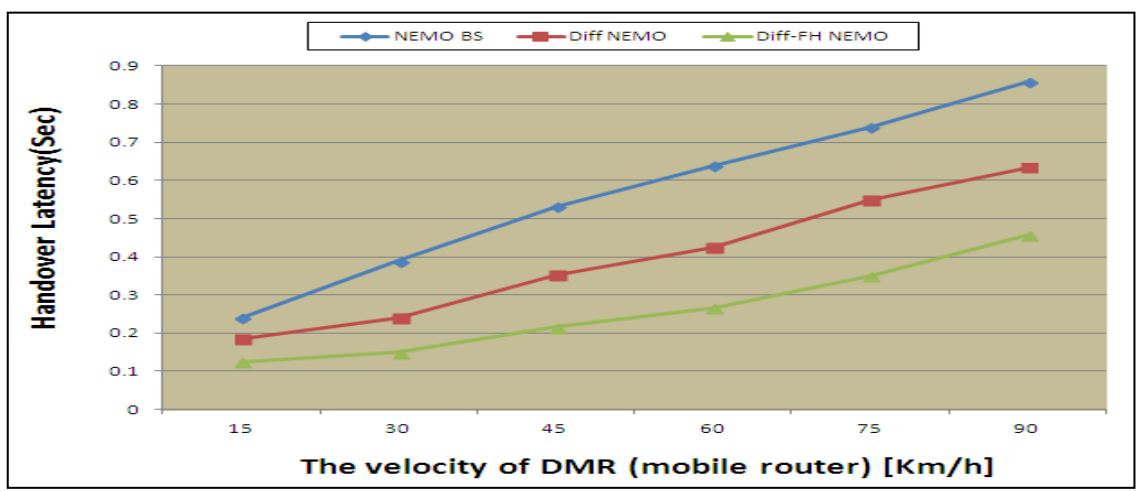

Figure 15. The handover latency versus speed

The impact of micro-movement (namely when handover process between AR1 and AR2) and macro movement (namely when handover process between AR2 and AR3) on the total packet delay for the standard NEMO BS protocol compared to the proposed schemes during the simulation time is illustrated in Figure 16. It can be seen from the graph the proposed scheme (Diff-FH NEMO) obtains less packet delay in micro mobility mode rather than macro mobility mode and outdoes all of (Diff NEMO) and the standard NEMO BS protocol.

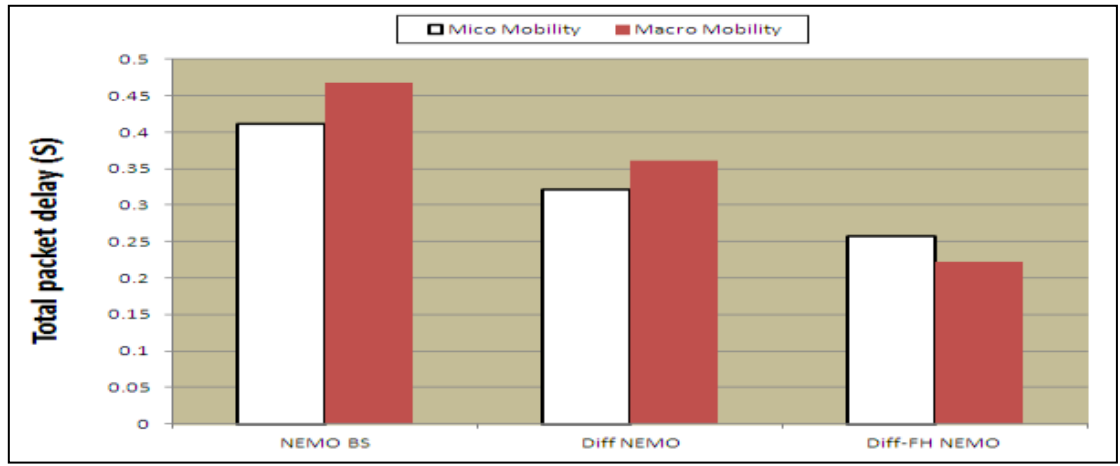

Figure 16. Total packet delay

## 6. CONCLUSIONS

In conclusion, nowadays the modern indispensable networking trend has been concentrated mostly on realizing all-IP mobile networks. In the information and communication technology, mobile networks perform a crucial role. The mobile networks can be sorted in terms of mobility support to host or network mobility. One of the main challenging issues for mobile networks is mobility management. Mobility management allows serving networks to find the attachment point of a mobile host to deliver information packets (i.e. location management) and to retain the communication of a mobile host as it continues to alter its attachment point (i.e. handover management). A well-designed mobile IP-based network should also not cause any significant deterioration of the service quality. Incorporation of QoS in IP-based networks has led to three distinct models namely, Differentiated Service (DiffServ), Integrated Services (IntServ) and Multiprotocol Label Switching (MPLS). Due to its scalability and simplicity advantages, the Differentiated Service appears to be the most commonly used in IP networks. However, this model was initially designed without mobility in-mind. Yet, it isn't completely adjusted to portable situations, particularly in a range that could passage as solo unit and join subjective spots within routing structure. In this paper, we have proposed a new scheme to enhance QoS in mobile network using DiffServ model. Furthermore, the proposed scheme takes the advantages not only with optimized communication route but also with the reduction of significant number of binding updates in the NEMO network. Network Simulation version 2 (NS2) tool has used to evaluate the performance of the proposed scheme. The simulation results reveal that the efficiency of the proposed scheme is better than NEMO BS protocol in terms of packet loss, packet forwarding rate and handover latency.

### ACKNOWLEDGEMENTS

The authors would like to thank the anonymous reviewers for their suggestions and valuable comments for improving the quality of this paper. This research was financially supported by Malaysian Ministry of Science Technology and Innovation (MOSTI) E-Science Fund, Project No. 01-01-08-SF0186, under the umbrella of Research Management Centre (RMC) at International Islamic University Malaysia.